\documentstyle[12pt]{article}

\def \be  {\begin{equation}}
\def \ee  {\end{equation}}
\def \beq  {\begin{equation}}
\def \eeq  {\end{equation}}
\def \ba  {\begin{eqnarray}}
\def \ea  {\end{eqnarray}}
\def \baa {\begin{eqnarray*}}
\def \eaa {\end{eqnarray*}}
\def \lab #1 {\label{#1}}
\newcommand\bqa {\begin{eqnarray}}
\newcommand\eqa {\end{eqnarray}}

\newcommand{\bear}{\begin{array}}
\newcommand{\enar}{\end{array}}

\font\cmss=cmss12 
\def\inbar{\,\vrule height1.5ex width.4pt depth0pt}
\def\IC{\relax\hbox{$\inbar\kern-.3em{\rm C}$}}
\def\IZ{\relax{\hbox{\cmss Z\kern-.4em Z}}}
\def\IR{{\hbox{{\rm I}\kern-.2em\hbox{\rm R}}}}

\def\IP{{\hbox{{\rm I}\kern-.2em\hbox{\rm P}}}}
\def\II{\hbox{{1}\kern-.25em\hbox{l}}}

\begin{document}

%\begin{titlepage}

\hfill\parbox{40mm}
{\begin{flushleft}  ITEP-TH-45/03
\end{flushleft}}

\vspace{1mm}

\centerline{\large \bf Spin Chains and Gauge/String Duality }

\vspace{7mm}

\centerline{\bf   A. Gorsky}

\centerline{\it Institute of Theoretical and Experimental Physics}
\centerline{\it B. Cheremushkinskaya 25, Moscow, 117259, Russia}

\vspace{1cm}
\centerline{\it{In  memory of Yan Kogan}}
\vspace{1cm}
%}
\centerline{\bf Abstract}

\vspace{5mm}

The  stringy picture behind the integrable spin chains
governing the  evolution equations in Yang-Mills theory is discussed.
It is shown that one-loop dilatation operator in N=4 theory
can be expressed in terms of two-point functions on 2d worldsheet.
Using the relation between Neumann integrable system and the spin chains
it is argued that the transition to the finite gauge theory coupling
implies the discretization of the worldsheet. We conjecture
that string bit model for the
discretized worldsheet corresponds to the representation of the integrable
spin chains in terms of the separated variables.

%\end{titlepage}

\vspace{2mm}

\section{Introduction}

The explicit realization of the string/gauge duality remains
the challenging problem during the last decades since the
early formulations \cite{polyakov}. The important
result \cite{maldacena} in N=4 SUSY YM case
is the
direct mapping between the modes of the closed string in
$AdS_5\times S_5$ background and the operators on the gauge theory side \cite{bmn,gkp,ft}.
The anomalous dimensions of the gauge theory operators are
identified with the energies of the corresponding stringy
excitations. However the problem is far from being completed
even in the most symmetric N=4 case.
The key problem is that there is no prediction for the
anomalous dimensions of the gauge theory operators at finite
coupling or equivalently  the proper formulation of the
string picture for the weak coupling
in the gauge theory is lacking. At the strong coupling semiclassical Nambu-Goto string picture
in $AdS$ type background
is valid yielding, for instance, explicit prediction for the
Wilson loop observable \cite{wilson}. However, at weak coupling
the problem is more involved since the string has to be considered
quantum mechanically and the whole spectrum of the stringy
modes has to be taken into account. Hence the
intermediate goal is to identify the closed string dynamics relevant
both for the
weak and strong coupling regimes and get the complete mapping
between generic multiparticle  conformal
operators and the stringy modes along the gauge/string duality
program.

Another approach to the calculation of the anomalous dimensions
of gauge theory operators deals with  open strings. It involves calculation
of the Wilson line observables for some specific contours.
The Wilson loop approach  in
QCD formulated in  \cite{polyakov} and  has been recently revisited in \cite{bgk}
with the emphasis to the underlying stringy picture. The calculations
of the Wilson lines at strong coupling matches precisely
with the closed string calculations \cite{kru} and
it was argued in \cite{bgk} using the mapping to topological
theory
that the relevant string at the weak coupling
seems to be tensionless in agreement with the earlier expectation.
In particular, the cusp anomaly which is a generating function
for the anomalous dimensions of the large spin twist two light-cone
operators \cite{km} has been related
at weak coupling to the disc amplitude in the two-dimensional
Yang-Mills theory admitting the stringy representation.

Generically operators  mix along the
RG evolution therefore to get the anomalous dimensions of the
multiplicatively renormalized operators one has to diagonalize
the dilatation operator which  can be represented
as the infinite or finite  dimensional matrix in the Hilbert space
of the gauge theory. It appears that the dilatation operator
discussed in \cite{dilatation}
at one loop level coincides with the Hamiltonian of finite
dimensional integrable systems.
Actually the phenomena of hidden integrability of the RG flows at large N
at least at one loop level seems to be quite general.
It has been found for low twist light-cone operators \cite{spins}
and arbitrary twist quasipartonic operators in SL(2,R) sector \cite{bgk}, operators
with large R charge in N=4 SYM theory \cite{mz} as well as for
evolution in the Regge limit \cite{regge}. Precisely, it appeared
that the dilatation operators at one loop level in these cases
can be identified with Hamiltonians of SL(2,R), SO(6) and SL(2,C)
spin chains respectively.
Recently the unifying integrable spin chain on $SU(2,2|4)$ supergroup
responsible for the evolution
of N=4 SUSY operators has been formulated \cite{beisert}.
The integrability of the many
body system implies the existence of   higher integrals
of motion which are in involution with the dilatation operator.
Hence the anomalous dimension of the operator involving N
constituents generically depends on (N-2) hidden quantum
numbers. Another important issue concerning the role of the nonlocal
integrals of motion has been recently raised \cite{wadia,polchinski,dolan}
(see also \cite{mikhailov} for the earlier discussion on
this point).

Within the gauge/string duality approach one has to find
out the proper place for these integrable systems.
It was argued in \cite{gkk} that the hidden
integrability could be the key for the stringy
description of the weak coupling limit in the gauge theory. Moreover
there are  very generic features of the integrable systems,
like existence of the huge number of
local and nonlocal integrals of motion,
appearance of the Riemann surfaces as the spectral curves,
the possibility to separate the variables e.t.c. which have
to be identified on the stringy side. The immediate idea for
the integrability to appear is the discretization of the worldsheet
which could amount to the integrable finite dimensional system.
Such approach has been discussed long time ago in the context
of the string bit model in the light cone gauge
\cite{thorn}.

In this paper we shall look for the stringy origin of these
integrable systems. First, we shall consider the different representations
of the cusp anomaly yielding the anomalous dimensions of the light-cone
operators with large SL(2,R) spin. It was argued in  \cite{bgk} that it allows the representation
in terms of the propagator of the particle on SL(2,R) group manifold
or as the disc partition function of SL(2,R) 2D YM theory. We shall
show that one more representation providing the worldsheet
description involves the two point function in the massive free
2d scalar theory. Since the cusp anomaly provides the anomalous dimensions for
large spin operators only we shall look also for the similar
two point representation in two dimensional
theory  for the arbitrary spin operators.  It appears
that it follows from the same scalar theory after the rotation
to the Rindler Hilbert space in the worldsheet theory.
Another argument involves the generalization of the particle motion
on SL(2,R) manifold or equivalently $AdS_3$ to  the  string
motion on the same manifold.

Apart from the cusp anomaly derivation of the
anomalous dimensions we shall try
to identify the origin of the spin chains. The natural
playground for the spin chains governing the evolution of light cone
operators appears to be
exactly solvable 2+1 dimensional gravity
\cite{witten3d}. For the negative cosmological constant
case it is described by the
SL(2,C) CS theory while for the flat space
one has ISO(2,1) gauge group. Moduli of the flat connections representing
the phase space of CS theory are mapped into the moduli
of the complex structures of the fixed time surfaces
in the Hamiltonian treatment of 3d gravity.
To get spin chains one has to add Wilson lines representing
point particles and consider the moduli space of flat connections
on the surface with punctures \cite{alekseev}. Then the monodromies
of the connection around the marked points can be related to the
local Lax operators of the spin chains.
Actually the relation of the statistical models with
knots formed by Wilson lines has been discussed
long time ago \cite{wittenknots}. To get  pair Hamiltonian
in higher spin chain we  use the Hamiltonian representation  of
dynamics of two-particles
coupled to the quantum gravity in three
dimensions. It appears that
the corresponding Hamiltonian coincides with the one governing the evolution
of two particle conformal operators.

This derivation concerns the one-loop anomalous dimension and
it is highly desirable to get the spin chains starting
from the strong coupling side. We shall make step in this direction
using the relation between two integrable finite dimensional
systems. Namely there is the mapping between the discretized
Neumann system and the stationary solutions to the XYZ spin chains \cite{veselov}.
Recently it was shown \cite{arut} that Neumann system describes the solution
to the equation of motion for the classical  string moving along $S_5$
part of the background. Using the relation with XYZ chain we shall
argue that the spin chains emerge if we take into account
correction to the strong coupling regime  amounting from the
discretization of the string worldsheet.

There is the natural procedure of the separation of variables
in the integrable systems so one could  discuss
its meaning in stringy terms and in general context of RG flows.
To get the geometrical insight one could
start with the classical spin chains whose solutions
to equation of motion are described by the
families of Riemann surfaces which
are invariants of RG flows. Effectively there is
one to one mapping between the  operators
and Riemann surfaces whose moduli are parameterized
by the anomalous dimensions of the operators and higher
integrals of motion \cite{gkk}.
We shall argue that the natural
stringy counterpart of the separated variables in the
spin chains is the string bit model in the light cone gauge.

The paper is organized as follows. In Section 2 we shall discuss
the representation of the anomalous dimensions
of large SL(2,R) spin operators
amounted from the cusp anomaly and the finite
spin case in terms of  two point functions
in the worldsheet massive free scalar theory
with a proper choice of the Hilbert space. In Section 3 we shall consider the
interpretation of  local Lax operators of spin chains  and
corresponding local Hamiltonian in terms
of Chern-Simons theory with noncompact group  and 3d quantum gravity.
The relation between Neumann system and spin chains is used
in Section 4 to argue that the finite coupling on the
gauge theory side implies the discretization of the
worldsheet.
Section 5 is devoted to the
possible stringy interpretation of the separation of variables procedure
wellknown in the integrability framework. Conclusion and
some open problems can be found  in Section 6.

\section{Cusp anomaly and two point correlators in 2D theory}

In this section we shall develop the 2d representation
for the anomalous dimensions of leading twist operators
on the light cone. Since the anomalous dimensions of the
4D operators can be extracted from their two-point correlators
it is desirable to get their derivation from 2D two-point
functions. To this aim we shall find the corresponding
worldsheet field theory in two dimensions.  One could distinguish
operators with large SL(2,R) spins which admit a kind
of quasiclassical description and the operators with generic spins.
We shall see that the transition from  large
to arbitrary spin corresponds to the proper choice
of the Hilbert space in two dimensional worldsheet theory.

Let us first reconsider the derivation of the anomalous
dimensions of the large spin operators from the cusp
anomaly. To this aim recall the perturbative calculation
of the Wilson line with cusp.
Cusp anomaly amounted from the integration of the gluon
propagator along two  lines. In the stringy calculation
these lines provide the boundary of the open string worldsheet.

To the lowest order in the coupling constant, the Wilson line expectation value
is given by
\be
W = <P \exp{i \oint_{\wedge} dx_\mu A^\mu(x)}>= 1 + \frac{(ig)^2}{2} t^a
t^a
\int_C dx_\mu\int_C dy_\nu \, D^{\mu\nu}(x - y) + O(g^4)
\, ,
\label{W-1-loop}
\ee
where $D_{\mu\nu}(x - y) \delta^{ab} = <0|T A^a_\mu(x)A^b_\nu(y)|0>$ is a
gluon propagator and $t^at^a=N_c$ is the Casimir operator in the adjoint
representation of the $SU(N_c)$. To calculate the cusp anomaly we choose the
integration contour $C$ with the single cusp and obtain
\be
W (v\!\cdot\!v')
=
1 - \frac{\alpha_s N_c}{\pi} \Big( w (v\!\cdot\!v') - w(1) \Big)
+
O(\alpha_s^2)
\, ,
\label{one-loop}
\ee
where $v_\mu$ and $v'_\mu$ are tangents to the integration contour in the
vicinity of the cusp, $v^2 = v'{}^2 = 1$, $v\!\cdot\! v' = \cosh\theta$,
$w(1) = w(v\!\cdot\!v) = w(v'\!\cdot\!v')$ and
\be
w (v\!\cdot\!v')
=
\int^0_{-\infty} d s \int_0^\infty d t \frac{v \!\cdot\! v'}{( v s - v' t )^2}
\, ,
\label{w}
\ee
with $s$ and $t$ being proper times. In higher orders in $\alpha_s$,
one takes into account that the Wilson loop possesses the property of non-abelian
exponentiation
\begin{equation}
\label{WilsonExponentiation}
\ln W = \sum_k { \frac{\alpha_s N_c}{\pi} }^k w_k \,,
\end{equation}
where the weights $w_k$ receive contribution from diagrams to the $k$-th order
in $\alpha_s$ with maximally nonabelian color structure, hence
$w_1=w(v\cdot v')-w(1)$.
The explicit integration amounts to
\be
\label{WinTermsOfChebyshev}
w(v \!\cdot\! v')
=
\theta\coth\theta \, \ln(\mu r_{\rm max})
\, ,
\ee
where $r_{\rm min}\sim 1/\mu$ and $r_{\rm max}$ are ultraviolet and infrared
cut-offs, respectively.

It was shown in \cite{bgk} that the one-loop cusp anomaly can be
represented as the transition amplitude of a particle on SL(2,R) group
manifold.
Indeed, one could perform a quantization procedure using four-dimensional
(Euclidean) polar coordinates $r^2=x_\mu^2$ and $v_\mu = x_\mu/r$.
Then, in the radial quantization the lowest order contribution to the cusp
anomaly takes the factorized form
\begin{equation}
w(v \!\cdot\! v')
= - \frac{1}{2}
(< -v' | \frac{1}{L^2} | v > + 1 ) \int {d \ln r}
\end{equation}
where
\be
L^2 = \frac{1}{8} l^{\mu\nu}l_{\mu\nu}  \qquad l_{\mu\nu} =
i ({x_\mu\partial_\nu-x_\nu\partial_\mu})
\ee
and $v$ denotes a point on the (hyper)sphere $SO(3,1)/SO(3)$ defined by
the unit vector $v_\mu$.  Thus at the
weak coupling, the cusp anomalous dimension is given by
\be
\Gamma_{ cusp} (\theta; \alpha_s)
=
- \frac{\alpha_s N_c}{2 \pi}
(< - v' | 1/L^2 | v >
-
< v| 1/{L}^2 | v >)
+
O(\alpha_s^2)
\ee

It is known that the dynamics of particle on the group
manifold can be related to the amplitudes in 2D YM
theory with the same group after a kind of T-duality transformation.
Therefore cusp anomaly can be interpreted equivalently as the disc partition
function of SL(2,R) YM theory integrated over the area $\tau$ \cite{bgk}
\be
{\mit\Gamma}_{\rm cusp} (\theta; \alpha_s)
=
-
\frac{\alpha_s N_c}{2 \pi}
\int_0^\infty d \tau
\Big(
{\cal Z} [U; 2 \tau] - {\cal Z} [\II; 2 \tau]
\Big)
\, .
\label{cusp=string}
\ee
where U is the holonomy around the disc boundary,
related to $\theta$ as $TrU(\theta)=2cosh\theta$.

To get the anomalous dimension of the large spin S operator from the cusp anomaly
one has to perform the following substitution
at large $\theta$ \cite{km}
\beq
e^{\theta} =iS
\label{cont}
\eeq
which means the analytic continuation with respect
to the cone variable if one considers the $\Pi$-shape
contour for the Wilson line , where
piece of the contour is close to the light cone.

We shall argue now that cusp anomaly can
be  formulated in the second quantized
2D worldsheet picture as two point function.
To this aim consider two dimensional worldsheet
field theory  with the equation of motion

\beq
(\partial_{t}^2 -\partial_{x}^2)\phi +m^2\phi=0
\eeq
whose solution has the following mode expansion

\beq
\phi(x,t)=\int \frac{d\beta}{2\pi} (a^{*}(\beta)e^{-im(xsinh\beta
-tcosh\beta} + a(\beta)e^{im(xsinh\beta
-tcosh\beta})
\eeq

It is convenient to introduce Rindler coordinates

\ba
x=rcosh\theta,\qquad t=rsinh\theta
\nonumber\\
-\infty < \theta < +\infty \qquad 0 < r <+\infty
\ea
in the space-time region $x>|t|>0$. Let us perform
the following Laplace transform with respect to the radial
coordinate

\beq
\lambda_{\theta}(\alpha)=\int dr
e^{imrsinh\alpha}(-\frac{1}{r}\partial_{\theta} +
imcosh\alpha)\phi(r,\theta)
\eeq
Then the commutation relation for  the Laplace transformed
field reads as
\beq
[\lambda(\alpha_1), \lambda(\alpha_2)]=i\hbar tanh(\alpha_1 -\alpha_2)/2
\eeq
and the Hilbert space is spanned by vectors $a(\beta_n)\dots
a(\beta_1)|vac>$ where vacuum state is defined as
\beq
a(\beta)|vac>=0 \qquad <vac|a^{+}(\beta)=0
\eeq
One can introduce two point function
\beq
F(\alpha_1-\alpha_2)= <vac|\lambda(\alpha_1)  \lambda(\alpha_2)|vac>
\eeq
and it appears that explicit calculation amounts to the
following answer \cite{luk}

\beq
F(\alpha -i\pi)= - \frac{1}{\pi}\alpha /2 coth(\alpha /2) + singular \quad terms
\eeq
The singular terms cancel
in the difference $ F(\alpha -i\pi) - F( 0)$ which coincides with the cusp anomaly in agreement
with the interpretation of \cite{bgk} in the first quantized picture.

However the cusp anomaly provides  the anomalous
dimensions only for large SL(2,R) spin S  operators  $\gamma_S\propto logS$ which
is large S asymptotic of $\psi(S)$ where $\psi(x)= \frac{dlog
\Gamma(x)}{dx}$. Hence the natural question concerns the modification
of two-point correlator which would yield the correct arbitrary spin
behaviour. The answer turns out to be remarkably simple; it is just necessary
to make the rotation in the Hilbert space of the theory from the
canonical Minkowski vacuum to the Rindler one attributed to the semiinfinite line $x>0, t=0$.
The crucial point is
that in the quantum gravity the Hilbert
space depends on the coordinate system; this property is responsible, for
instance, for the Unruch effect for the accelerating observer. The proper
representation of this space goes as follows. Consider
the expansion of the solution to the Klein-Gordon equation
\beq
\pi \phi(r,\theta)=\int dk K_{ik}(r)[b_k e^{-ik\theta} +  b_k^{+}
e^{ik\theta}]
\eeq
where $K_{ik}(r)$ are Macdonald functions and oscillators are
related with functions $\lambda(\alpha)$ as

\beq
\lambda(\alpha +i\pi/2)= i \int dk \frac{b_k e^{ik\alpha}}{sinh \pi k}
\eeq
Hence new oscillators obey the commutation relations
\beq
[b_k, b_{k'}]= sinh(\pi k)\delta(k+k')
\eeq
and Rindler vacuum is defined as
\beq
b_k|0>= <0|b_{-k}=0,\qquad k>0
\eeq
Note that  Minkowski vacuum
can be interpreted as a kind of coherent state in the
pair of vacua in two Rindler wedges
\be
|0>_{Min} =\prod _{k}exp(e^{-2\pi \omega_k}
a^{+}_{R_{-},k}a^{+}_{R_{+},k})|0>_{Rind}
\ee
where $a^{+}_{R_{\pm},k}$ are the creation operators in left(right)
Rindler wedges.
Two point correlator calculated in the Rindler vacuum equals \cite{luk}
\beq
<0|\lambda(\alpha_1)  \lambda(\alpha_2)|0>
=\pi \psi(1/2 +i\alpha/2\pi) +const
\eeq
After substitution $\frac{i\alpha}{2\pi}= J$ similar to (\ref{cont}) , one gets the correct anomalous
dimension of twist two correlators. Hence the complete one-loop
dilatation operator in N=4 SYM theory found in \cite{beisert}
\beq
D=\sum_{k}H_{k,k=1}, \quad H_{k,k+1}= \sum_{j}h(j)P_{k,k+1|j}
\eeq
where h(j) are harmonic numbers and $P_{k,k+1|j}$ projects two body Hilbert space onto
fixed representation of $PSU(2,2|4)$, can be expressed in terms of two point correlators in 2d theory.

Since the cusp  anomaly is defined by the Wilson lines
"correlator" in 4D theory one could wonder if 2D
variable $\lambda(x)$ can be also formulated in terms of some 2D Wilson
lines. To this aim we can
artificially represent
Klein- Gordon equation as the zero curvature condition
for SL(2,R) connection.
To this aim let us start with Liouville model
\beq
(\partial_t^2 -\partial_x^2)\phi + \frac{m^2}{b}e^{b\phi}=0
\eeq
allowing
zero curvature condition for SL(2,R) valued connection $A_{\theta},A_{r}$.
Then one could introduce the Wilson line or monodromy
\beq
{\bf{T}}^{\theta}(\alpha) \propto e^{imRsinh\alpha \sigma_3}Pexp(\int dr
A_{r}(r,\theta,\alpha))
\eeq
where R is cutoff, which defines $\lambda_{Liov}(x)$ via relation
\beq
\lambda_{Liov}(\alpha) =-ilnT_{11}(\alpha)
\eeq
The latter reduces to Klein-Gordon field $\lambda_{KG}(\alpha)$ in the weak coupling limit
$b\rightarrow 0$ \cite{luk}
\beq
\lambda_{Liouv}(\alpha) \rightarrow \frac{b}{4} \lambda_{KG}(\alpha)
\eeq

Leu us make one more comment.
We have seen that the cusp anomaly has the interpretation of the
transition amplitude for the particle on $AdS_3$ manifold. The natural
expectation is that transition amplitude for a string on $AdS_3$
emerges for the operators with arbitrary spins. To trace out the proper
stringy degrees of freedom let us represent the one loop dilatation operator
relevant for the two particle operator in the SL(2,R) as sum
\beq
H_{k,k+1}=\sum_{l} \frac{2l+1}{l(l+1) +(J_k +J_{k+1})^2}
\label{sum}
\eeq
where $J_k,J_{k+1}$ represent two SL(2,R) spins.
The sum looks similar to $L_0^{-1}$ if one assume
that pair Casimir contribution
can be attributed to the string ends.
However more precise interpretation in terms of the
propagator  in the string field theory has to be found.

\section{Spin chains from gravity}

It was shown in \cite{bgk} that the cusp anomaly
can be interpreted in terms of the observables in two
dimensional BF theory with SL(2,R) group which is
equivalent to Jackiw-Teitelboim dilaton gravity. Here we shall
discuss some generalization of this observation and
shall show how the main ingredients of spin chain
machinery, namely Lax operator as well pair Hamiltonian
derived from the fundamental R matrix can be defined
in the gravity framework.

We shall exploit the general approach
for generation of the integrable models from the
special observables in Chern-Simons theory. The key
idea is that from the multiple Wilson line observables
one derives the transfer matrix of the vertex like
integrable model . We shall consider
Chern-Simons theory with noncompact group related to gravity
\cite{wittencs}.
Such theories appear in the context of three dimensional
gravity at least in two ways. First, $AdS_3$ gravity
is described by CS theory with SL(2,C) gauge group.
On the other hand string theory on $AdS_3$ with
background NS form
is equivalent to WZW model with SL(2,R) group. This
theory is one which SL(2,R) Chern-Simons reduces to
on the boundary according to the standard CS-WZW relation.
The noncompact SL(2,R) group plays the role of
collinear conformal group governing the renormalization
of the light cone operators.

We shall first remind how the fundamental and auxiliary
transfer matrices of
spin chain  with generic spins emerges  from the R-matrix formalism
(see, for instance, \cite{faddeev}) and then argue that
the same objects follow from three dimensional gravity.
The starting point is the universal $\cal{R}$ matrix which
amounts to the concrete Lax operators when taken in
the evaluation representation
\beq
L_{n,m}(\lambda -\mu)= (\rho(n,\lambda) \otimes
\rho(m,\mu)){\cal{R}}=R_{n,m}(\lambda-\mu)
\eeq
where representation $\rho(n,\lambda)$ can be considered as the
representation of the Yangian algebra for XXX model and
quantum affine algebra for XXZ model. These algebras
play the role of the algebras of observables for the
corresponding dynamical systems. The relevant
projection of the Yang-Baxter equation
on the triple of representations looks as
\beq
R_{s_1,s_2}(\lambda-\mu)R_{1/2,s_2}(\sigma -\mu)R_{1/2,s_1}(\sigma
-\lambda)=R_{s_1,s_2}(\lambda-\sigma)R_{1/2,s_2}(\sigma -\mu)R_{s_2,s_1}(\mu
-\lambda)
\label{YB}
\eeq
In what follows we shall
be interested in the case when $n=m=s$ and the corresponding Lax
operator $L_{F}$ is called fundamental.

The solution to the Yang-Baxter equation is looked for in the following form
\beq
R_{s_1,s_2}=P_{s_1,s_2}r((S,T),\lambda)
\eeq
where $P_{s_1,s_2}$ is a permutation between two (2s+1) dimensional spaces
and $(S,T)$ is a pairwise Casimir operator.
The equation (\ref{YB}) reads as
\beq
(\lambda S_a +(T\times S)_a )r(\lambda)=r(\lambda)(\lambda T_a +
(T\times S)_a)
\eeq
which can be reduced to the functional equation
\beq
(\lambda +iJ)r(\lambda,J-1)=r(\lambda,J)(-\lambda+iJ)
\eeq
with solution
\beq
r(J,\lambda)=\frac{\Gamma(J+1+i\lambda)}{\Gamma(J+1-i\lambda)}
\eeq

From the product of local
fundamental Lax operators $L_{i,F}$ we can get the fundamental transfer matrix
$T_{F}(\lambda)$ commuting as
\beq
[T_{F}(\lambda),T_{F}(\mu)]=0
\eeq
Moreover due to the Yang-Baxter equation fundamental transfer matrix $T_{F}$
commutes with the auxiliary one $T_{a}$ obtained from $R_{1/2,s}$ representations
\beq
[T_{F}(\lambda),T_{a}(\mu)]=0
\eeq
The Hamiltonian of the spin chain describing the anomalous dimensions
follows from the fundamental transfer matrix
\beq
H=i\frac{d}{d\lambda} lnT_{F}(\lambda)_{\lambda=0}
\eeq

Let us discuss now how the fundamental and auxiliary transfer matrices
emerges from Chern-Simons theory. Let us start with derivation of the
auxiliary transfer matrix \cite{alekseev}
and  consider SL(2,R) CS theory with  Wilson lines added.
The monodromies around the marked points are
\beq
M_i=P exp\int_{\Gamma_i}A
\eeq
where $\Gamma_i$ is the contour around i-th marked point.
Introduce the monodromy of the flat connection with the spectral parameter
\beq
M_i(\lambda)= M_i +\lambda I
\eeq
where I is the unit matrix.
This matrix is gauge equivalent to the local
L operator for  the XXZ spin chain
\beq
M_i=K_{i-1}(L_{+}(i) + \lambda L_{-}(i))K_{i}^{-1}
\eeq
where we introduced matrix $L_i$ obeying the
standard quadratic  R matrix relations with
the decomposition
\beq
L_i= L_{+}(i)L_{-}(i)^{-1}
\eeq
and
\beq
K_{i}=L_{-}(1) \dots L_{-}(i-1)
\eeq
Therefore
\beq
M_i=K_i L_i K_{i}^{-1}
\eeq
and the traces of monodromy matrix for $M_i(\lambda)$ and $L_i(\lambda)$
coincide \cite{alekseev}  amounting to
the desired monodromy matrix of the spin chain indeed.
Note that the spin chain arises on the string worldsheet
not target space in according with the expectations. Since
the anisotropy parameter depends on the level k
in the standard way quasiclassical limit $k \rightarrow \infty$
yields the XXX chain.
Let us emphasize that the representation of the spin chains
as the perturbed Chern-Simons theory
provides the desired stringy
picture for the compact group corresponding to the operators
involving $S_5$ geometry.
Indeed
the Chern-Simons theory with compact group has been proven to
have stringy reformulation \cite{wittencs} where
the level k defines the open string coupling constant.

We have shown above how the (1/2,s) Lax operator can
be derived from the Chern-Simons theory. Let us turn now to the
derivation of the spin chain Hamiltonian amounted
from the fundamental (s,s) R matrix. First, we shall
identify pair Hamiltonian from the point of view
of 3d gravity described by the CS action. To this aim
consider system of two particles in 3D gravity. It is
known that particle in three dimensions creates the
cone singularity around therefore we could expect the
Hamiltonian to describe the test particle moving on the
cone of aperture determined by the total energy.

It was argued in \cite{seminara} that in the second order ADM
formulation of gravity in three dimensions
coupled to the point particles the Hamiltonian
constraint in the maximally slicing gauge can be
presented as the Liouville equation with sources
\beq
\Delta \phi=-e^{\phi} -4\pi \sum_n \delta(z-z_n)(\mu_n-1) + 4\pi \sum_a \delta(z-z_a)
\eeq
where $\mu_n$ are the particle masses , $z_n$-particle positions
and $z_a$ are so called apparent singularities.

It can be shown that there are no apparent singularities
in the case of two particles and
the motion of the particles appears to be Hamiltonian with
\beq
H=lnPz +log\bar{P}\bar{z}
\eeq
where z is the relative coordinate of the particle
and P is the conjugate momentum.
The Hamiltonian above classically coincides with the
spin chain Hamiltonian with two sites.

In the many particle case the apparent singularities
emerge and the motion of particles is Hamiltonian
with \cite{seminara}
\beq
H=\frac{1}{2\pi} \frac{\partial S}{\partial \mu}
\eeq
where S is the Liouville action calculated on the solution, and $\mu$
is the total mass of the system. Its relation with the spin chains
Hamiltonians deserves further investigation.

\section{Spin chains from strong coupling limit}

Let us explain how spin chains emerge  starting from
the strong coupling limit. Recall that in
this regime the string can be considered classically
and the classical string energy yields the anomalous
dimensions of the corresponding gauge theory operator
which scales as $\sqrt {\lambda}$, where $\lambda=g^2 N$, at strong coupling.
The particular solutions to the classical equations
of motion of string in $AdS_5\times S_5$ background
correspond to some N=4 gauge theory operators.  There
are some useful anzatzs parameterizing the classical string
motions. In particular the following anzatz has been considered
in \cite{arut} parameterizing generic motion on $S_5$
\beq
X_1 +i X_2=x_1(\sigma)e^{\omega_1 \tau} \qquad
X_3 +i X_4=x_2(\sigma)e^{\omega_2 \tau} \qquad
X_5 +i X_6=x_3(\sigma)e^{\omega_3 \tau}
\eeq
where $X_i$ correspond to coordinates of $S_5$ part of geometry.
It corresponds to string rotating in $S_5$ with three angular
momenta $\omega_i,i=1,2,3$. If one substitutes this anzatz into
the equation of motion then
it reduces to the Neumann finite dimensional integrable  system whose
energy yields  the anomalous dimension of operators with
nonvanishing R charge \cite{arut}.

We know that spin chains should emerge if we start to take into account the
quantum corrections to the classical solutions. Let us demonstrate that
they amount from the discretization of the string worldsheet. The key
point is that there exists mapping between the discretized Neumann system
and XYZ spin chains \cite{veselov}. It can be formulated as follows;
the stationary solutions to the XYZ classical equations of motions
coincide with the discrete time evolution in the Neumann system.
Namely, let us consider the stationary equation for the XYZ chain
with the classical Hamiltonian
\beq
H= \sum_{i} (S_i,JS_{i+1})
\eeq

\beq
S_{i+1} +S_{i-1}= 2(S_{i-1},J^{-1}S_i)(J^{-1}S_i)^{-2}J^{-1}S_i
\eeq
The solutions to this equation look as \cite{veselov}
\beq
S_k^{n} =  \beta_k J_{n} \frac{\theta(k \vec{U}+\vec{\rho})}{\theta(\vec{\rho)}}
\eeq
where $ \vec{U}$ is the vector of b periods of the
abelian differential of the third kind $\Omega$
normalized by conditions
$\oint_{a_i}\Omega =0$ and $\beta_i$ is determined via
the normalization condition $S_i^2=1$.
Differentials are defined on the  Riemann surface which is
the spectral curve
of the Neumann dynamical system.

In the discrete time evolution it is implied that we have
dynamical mapping corresponding to the jumps between
the points on the spectral curve instead of the continuous motion
in the canonical Neumann system. The value of the degree of freedom
$x_i$ at k-th time interval is identified with the value of the i-th
spin variable on the k-th site of the chain $S^i_k$. From the stringy point of
view the time variable in the Neumann system corresponds to the coordinate along the
closed string hence time discretization corresponds to the discretization
of the string worldsheet and string is essentially substituted by the
set of spin degrees of freedom $S^i_k$.

Several comments are in order here. First, one could wonder how the
genera of the spectral curves match for two models. Naively,
the genus of  the Neumann spectral curve is two while the genus
of the spin chain curve is higher depending on the number of cites.
However one has to take into account that the stationary solution
to the chain equation of motion enters the correspondence.
For the stationary solution the genus reduces and exactly coincides
for two models. Similar reduction for the genus on the stationary
solutions has been observed recently in \cite{deboer} in the
different context. Another point to be mentioned is that Hamiltonian
of the discretized Neumann system can be identified with the nonleading
integral of motion in the spin chain
\beq
H_3=(JS_n)^2 +(JS_{n+1})^2 -(S_n,JS_{n+1})^2
\eeq
and reduces to the standard Neumann Hamiltonian in the
continuum limit.

Finally, the number of the cites in the spin chain has to be found.
To this aim one has to look at the solution to the classical
equations of motion and defines the period in the discrete  k variable
which is determined by U. The direct analysis shows that
the length of the chain is defined by R-charge for SO(6) chain.
In a similar manner
the string with discretized worldsheet  in SL(2,R) sector amounts
into  XXX chain with length proportional to the number of partons involved.
Hence we see that spin chains are related with the
semiclassical string motion  after worldsheet discretization. The
number of the sites in the chain can be determined by the classical
string motion.

\section{ Separation of variables in integrable systems and  string bit model }

Weak coupling regime in gauge theory corresponds to the
strong coupling regime in $\sigma$-model. One approach to deal
with strong coupling theories is to discretize the worldsheet
hence this procedure is natural for the problem under
consideration too. In the previous section we have seen that
discretization emerges naturally in the integrability
approach for the finite coupling constant.
Corresponding model with the discretized
worldsheet is known as string bit model \cite{thorn}. It is
defined for light-cone string and the number of bits corresponds
to the decomposition of the total momentum $P_{+}$ into fractions.
We shall discuss the emergence of the string bit model from the
perturbative calculations and conjecture the
identification of  the single bit as
the separated variable in the spin chains.

Let us first discuss the classical picture. The string bit model implies
that the phase space of the string can be represented
as $M^n/S^n$
for n bit case, where M
is  target space and $S^n$
is permutation group for n elements. For instance, in the plane wave geometry number of bits
coincides with the R-charge J \cite{verlinde} and
for SL(2,R) sector coincides with the number of partons
involved into the multiparticle operators. The crucial point is that
classical motion of bits has to be highly coherent to provide
the correct worldsheet of the whole string. Note that in spite of
the interaction between bits each of them has the similar classical motion.

Let us compare the general features of the string bit model with
the separation of variable procedure in the integrable systems.
Separation of
variables in any integrable system with n degrees of freedom
is interpreted as the canonical transformation
bringing the phase space into the form $M^n/S^n$
with two-dimensional complex manifold M depending on the model
under consideration. More detailed consideration shows
that the symmetric power must be substituted by the
Hilbert scheme $Hilb^{n}M$ \cite{duality}. The desired
"coherence" of classical motion looks as follows. One can
define the spectral curve for the integrable system
\beq
det(T(\lambda)- \eta)=0
\eeq
where $T(\lambda)$ is the transfer matrix acting in the auxiliary
space whose explicit form depends on the model. Then n separated
variables move classically along the higher genus Riemann surface
similar to motion of n bits which form the Riemann surface
as the worldsheet. Note that in brane language the separation
of variables procedure corresponds to the realization
of the system in terms of D0 branes \cite{duality}.

Since we are looking for the spectrum of the anomalous   dimensions
the quantum spectrum of spin chains is of the main interest.
At the quantum level the equation
of the spectral curve becomes the operator acting on the
function Q(x)  depending on the single separated variable.
This function Q(x) is known as the Baxter function in the context
of the integrable systems.
Then the
wave function roughly becomes the product of n Q functions depending on
separated variables $x_i$. This approach   has a lot in common with the transition from
the first to second quantization. Namely, the dispersion
relation of the free particle $p^2- m^2=0$ becomes
the equation on the field operator $(-\partial ^2 -m^2)\Phi(x)=0$.
When we consider the transition to the string bit model
the classical spectral curve of the spin chain plays the role
of the  dispersion relation  while the Baxter equation
corresponds to a kind of field equation.
We  conjecture that solution to the Baxter equation Q(x)
corresponds to the "wave function" of the single bit.
If single bit function Q(x) is presented in
polynomial form then the roots of the polynomial
obey the Bethe anzatz equations.

Let us make a few comments relevant for the spin chain systems
under discussion. The Baxter equation for
SL(2,R) sector looks as follows

\beq
trT_{F}(\lambda)Q(\lambda)=\Delta_{+}(\lambda)Q(\lambda +i) +
\Delta_{-}(\lambda)Q(\lambda -i)
\eeq
where $\Delta_{+},\Delta_{-}$ are elements $T_{11}$ and $T_{22}$
of the monodromy matrix, respectively.
It is important that  spectrum of  pair Hamiltonian $H_1=\sum H_{k,k+1}$ where $
H_{k,k+1}=\psi(J_{k,k+1})-\psi(2s)$
amounting to the spectrum of anomalous dimensions can be expressed entirely in terms
of the single Baxter function $Q(\lambda)$
\beq
E(H_n)=i\frac{d}{d\lambda} log \frac{Q(\lambda+i)}{Q(\lambda-i)}_{\lambda=0}
\eeq
Note that one more important representation for the pair Hamiltonian
involves integral operator \cite{derkachov}
\beq
H_{k,k+1} \Psi(x_k,x_{k+1})= \int_{0}^{1} dz \frac{(1-z)^{2s-1}}{z}
[\Psi((1-z)x_{k} +zx_{k+1},x_{k+1}) + \Psi(x_k,(1-z)x_{k+1} +zx_{k}) - \Psi(x_{k},x_{k+1})]
\eeq
It is this form of the Hamiltonian which emerges from the
direct calculations of the Feynmann diagrams. Let us remind that
in SL(2,R) sector $x_i$ are the fractions of the light-cone
momentum$P_{+}$.

One more comment is in order.
In the  general framework $Q(\lambda)$ can be considered as the
eigenfunction of the Baxter operator $\hat{Q}(\lambda)$ which commutes
with the canonical transfer matrix and therefore has common set of
eigenfunctions. Integral representation for the Baxter operator
looks as follows
\beq
\hat{Q}(\lambda)\Psi(x)= \prod_{k=1}^{n} \Gamma(\lambda,s)
\int dz_{k} z_{k}^{s-\lambda} (1-z)^{s+\lambda} \Psi(z_k x_k +(1-z_k)x_{k-1})
\eeq
where
\beq
\Gamma(\lambda,s) =\frac{\Gamma(2s)}{\Gamma(s+\lambda)\Gamma(s-\lambda)}
\eeq
The role of the Baxter operator in the context of representations
of the Virasoro algebra has been discussed in \cite{blz}.

The number of string bits can be counted in the quasiclassical limit.
The types of solutions to the classical equations of motion correspond
to the types of the integrable models. For instance motions
of bosonic string in $S_5$ amount to SO(6) chains while motion
in $AdS_3$ part of $AdS_5$ geometry is governed by SL(2,R) chains.
To some extend classes of string motions correspond to the
nonlocal operators on the gauge side which serves as the generating
functions for the local operators. Upon choice of the
class of the solutions to the string equations of motion that is
the type of the integrable system the number of degree of freedom
in the integrable system is fixed. On the stringy side
it corresponds to the number of string pieces reaching the boundary
of the background manifold. For instance, to get the operators of the type
$D^{n_1}\Psi...D^{n_k}\Psi$ on the cone
corresponding to the chain with k cites one has to consider the
string configuration with k "arms" \cite{bgk}.

As was observed \cite{mz} the  Bethe state without zeros
corresponds to the ground state of string, number
of zeros in the solution corresponds to the
number of different exited oscillators while
the number of complex zeros with fixed real part
corresponds to the number of exited oscillators
on the fixed level.
The number
of string bits corresponds to the number of degrees of freedom
in the SL(2,R) spin chain that is the number of fields involved
into the  conformal operator. Hence we have two bit model for the simplest
twist two operators and solution to the Baxter equation
corresponds to the single bit wave function. The next nontrivial
three bit string example involves the nontrivial torus topology of the spectral curve
at the classical level and corresponds to the nontrivial solitonic
like states at the quantum level which follows from
the effective creation of the "two bit bound state". Generically
the spectral curve for SL(2,R) sector is hyperelliptic while
in SO(6) sector it is multiple cover of the genus two curve.
The quasiclassical quantization suggested in \cite{gkk}
implies the hidden S duality in the spectrum of anomalous
dimensions which still has to be elaborated.

\section{Discussion}

In this paper we discussed the origin of the spin chains emerging in the context
of RG properties of Yang-Mills operators. We have shown how the cusp anomaly
which is generating function for the anomalous dimensions of large SL(2,R)
spin operators
as well as its proper finite spin generalization  can be derived
from two-point functions in 2D worldsheet free massive scalar theory.
We have also
demonstrated how the Lax operators and pairwise Hamiltonian of spin chains
emerge from 3D gravity. Elements of fundamental R matrix
projected to the fixed spin  can be expressed in terms of  correlators
in the worldsheet theory and it seems that proper information about
the complete dilatation operator is encoded in the universal R matrix.
It would be interesting to make contact with spin network states in
the loop gravity where quantization of the length \cite{rovelli} emerges
similar the quantization of the cone variable we observed for the
SL(2,R) type operators. The questions concerning the interpretation
of the issues above in terms of open/closed duality and twistor picture
for the anomalous dimensions shall be discussed elsewhere \cite{gg}.

Using the relation between the stationary solutions to
the equations of motion in the spin chains
and discretized Neumann system we have shown that transition to the
finite coupling in the gauge theory implies the discretization of
the worldsheet. The precise pattern of the discretization can be read off
from the explicit solution to the stringy equation
of motion in $AdS_5\times S_5$
background. The periodicity condition for the worldsheet coordinate
amounts to the fixed number of the "string bits". Generically the number of
string bits depends on the quantum numbers of the string with respect to the
isometry of background as well as coupling constant. For large spins
we have seen the agreement with the number of the cites in the spin chains
followed from the explicit one loop calculations.

We have also conjectured that the separation
of variables in the spin chain has the transition to the
string bit model as the stringy counterpart. The number of
spin cites coincides with the number of string bits indeed.
The crucial object along this line of reasoning is the
Baxter equation which is the quantization of the spectral curve
equation and its solution which essentially yields the spectrum
of the quantum model. We believe that further clarification
of the stringy role of the Baxter operator which was proved to
be essential ingredient in the
representations of the Virasoro algebra and its
discretization is very important. It is also desirable
to get  precise stringy picture of the quantization
conditions developed in \cite{gkk} as well as mapping
to the corner transfer matrix formalism.

I would like to thank A. Belitsky, A. Gerasimov and G. Korchemsky
for the useful discussions. The work was partially supported
by grant INTAS-00-334.

\end{document}